\documentclass[fleqn,10pt]{wlscirep}
\usepackage[utf8]{inputenc}
\usepackage[T1]{fontenc}
\usepackage{amsmath}

\title{X-ray Fokker--Planck equation for paraxial imaging}

\author[1,*]{David M. Paganin}
\author[1,2]{Kaye S. Morgan}
\affil[1]{School of Physics and Astronomy, Monash University, Clayton, Victoria, 3800, Australia}
\affil[2]{Chair of Biomedical Physics, Department of Physics, Munich School of Bioengineering, and Institute of Advanced Study, Technische Universit\"{a}t M\"{u}nchen, 85748, Garching, Germany}

\affil[*]{David.Paganin@monash.edu}

\begin{abstract}
The Fokker--Planck Equation can be used in a partially-coherent imaging  context to model the evolution of the intensity of a paraxial x-ray wave field with propagation. This forms a natural generalisation of the transport-of-intensity equation. The x-ray Fokker--Planck equation can simultaneously account for both propagation-based phase contrast, and the diffusive effects of sample-induced small-angle x-ray scattering, when forming an x-ray image of a thin sample. Two derivations are given for the Fokker--Planck equation associated with x-ray imaging, together with a Kramers--Moyal generalisation thereof. Both equations are underpinned by the concept of unresolved speckle due to unresolved sample micro-structure.  These equations may be applied to the forward problem of modelling image formation in the presence of both coherent and diffusive energy transport.  They may also be used to formulate associated inverse problems of retrieving the phase shifts due to a sample placed in an x-ray beam, together with the diffusive properties of the sample. The domain of applicability for the Fokker--Planck and Kramers--Moyal equations for paraxial imaging is at least as broad as that of the transport-of-intensity equation which they generalise, hence the technique is also expected to be useful for paraxial imaging using visible light, electrons and neutrons.
\end{abstract}
\begin{document}

\flushbottom
\maketitle

\thispagestyle{empty}

\section{Introduction} 

\begin{figure}
\centering
\includegraphics[width=470pt]{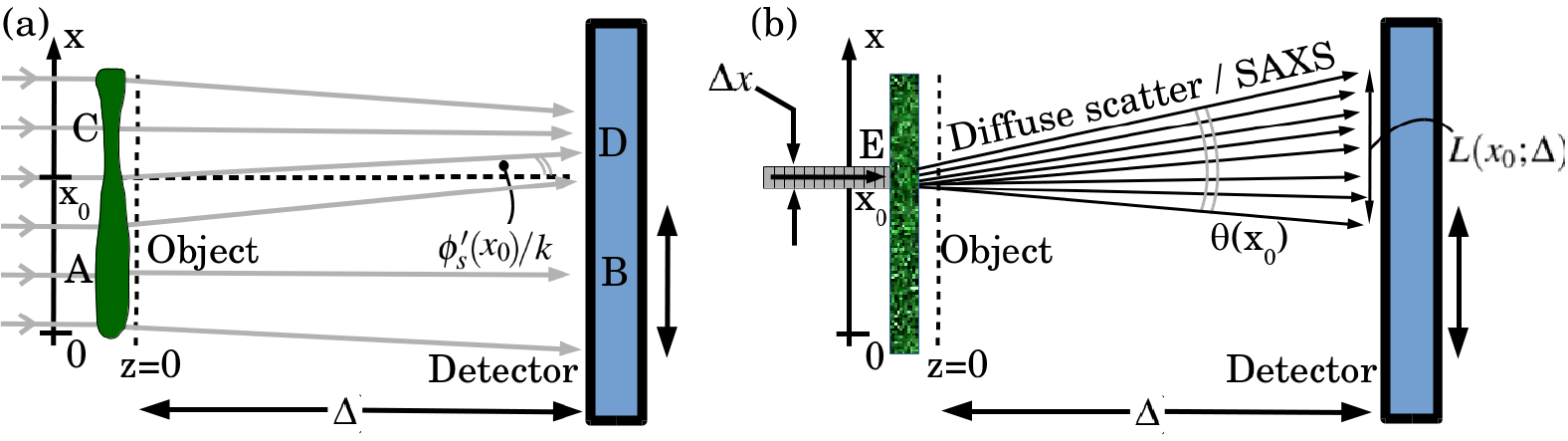}
\caption{(a) Specular refraction by a thin object and associated propagation-based phase contrast (local concentration and rarefaction of photon energy density) associated with coherent energy transport downstream of the object. (b) Diffuse scattering by a thin object and associated propagation-based blurring associated with diffusive energy transport downstream of the object.  Paraxial coherent energy transport may be modelled using the transport-of-intensity equation (see Eq.~(\ref{eq:TIE})), while paraxial diffusive energy transport may be modelled using the diffusion equation (see Eq.~(\ref{eq:DiffusionEquation})).  The Fokker--Planck equation can simultaneously model both effects, for small $\Delta$---see Eq.~(\ref{eq:FPEin2D}), together with the Kramers--Moyal generalisation in Eq.~(\ref{eq:KramersMoyal6}).}  
\label{fig:Samples}
\end{figure}

Suppose that coherent visible-light plane waves normally illuminate a thin lens made of glass that is slightly cloudy.  For such a lens, a large percentage of the light will be coherently focused or defocused upon propagation beyond the lens.  Conversely, a small percentage of the light will be diffusely scattered by the random cloudy inclusions within the glass.  In the hard-x-ray analogue of this experiment, we replace the glass lens with a thin semi-transparent sample whose large-scale projected structure locally resembles an aberrated converging or diverging lens (see Fig.~\ref{fig:Samples}(a)), and replace the cloudy inclusions with unresolved random micro-structure within the volume of the sample (Fig.~\ref{fig:Samples}(b)). For hard x rays we will often have a situation where both the coherent and diffuse scatter are narrowly peaked in the forward direction. With reference to Fig.~\ref{fig:Samples}(a), we consider the ``near field'' regime where the object-to-detector distance $\Delta$ is significantly smaller than the focal length of the ``local lenses'' of which the sample is considered to be comprised: see e.g.~the ``local diverging lens'' at $A$, and the ``local converging lens'' at $C$, the nominal focal length $f$ for each lens being much larger than $\Delta$.  This is the key situation, of {\em combined coherent and diffusive energy transport in a near-field paraxial imaging setting}, that we wish to consider in the present paper.  While we focus on the case of hard x rays, the methods considered here will also be applicable to paraxial imaging using visible light, electrons, neutrons etc.  

We make the simplifying assumptions that: (i) magnetic effects such as x-ray circular dichroism in the sample can be ignored; (ii) the sample is non-crystalline, allowing x-ray Bragg-diffraction effects to be ignored; (iii) the sample is sufficiently thin that dynamical x-ray diffraction effects, within the volume of the sample, can also be ignored; (iv) all polarisation-sensitive effects can be ignored.  Many well-known means exist, for the quantitative modelling of an x-ray beam interacting with such a sample and subsequently forming a paraxial image as it propagates through space\cite{RussoBook,XRayImagingBook1,XRayImagingBook2}.  One of the most often-employed, and most elementary, means is to first use the projection approximation to quantify the passage of an incident coherent scalar x-ray beam through the sample, and then use the Fresnel diffraction formalism to propagate the resulting exit-surface wave field to the surface of a detector\cite{RussoBook,Paganin2006}.  The projection approximation assumes that: (i) the radiation is of sufficiently high energy that individual rays within the sample (more precisely, streamlines of the energy-flow vector within a sample) are unchanged {\em within the volume of the said sample} in comparison to the streamlines that would have existed in the absence of the sample, and (ii) both phase and amplitude shifts are accumulated continuously and locally, as one traverses any such ray from the entrance surface of the sample to the exit surface---see e.g. Sec.~2.2 of Paganin\cite{Paganin2006}.  The Fresnel approximation, which is a consequence of the paraxial approximation, assumes that spherical Huygens-type wavelets can be approximated as having parabolic wave fronts.  If the sample-to-detector propagation distance is sufficiently small, and if diffuse scattering is negligible, Fresnel diffraction implies that the transport-of-intensity equation\cite{Teague1983} (TIE) may be used to model the evolution of image intensity with respect to propagation distance downstream of the sample, for a paraxial quasi-monochromatic beam.  The TIE, which is the continuity equation for paraxial wave fields,  expresses local energy conservation\cite{PaganinPelliccia2019}.  It forms a basis for several imaging techniques in coherent x-ray optics\cite{RussoBook,paganin2002,Paganin2006}. 
Like the formalism of Fresnel diffraction upon which it is based, the TIE models what may be called ``coherent energy transport'' downstream of a thin sample.  That is, at each point in the space downstream of the sample, there is {\em one} time-independent energy-flow vector (Poynting vector proportional to the intensity multiplied by the gradient of the wave-field phase\cite{GreenWolf1953,NietoVesperinasBook, BerryOpticalCurrents}).  However, if there is unresolved spatially random micro-structure within the sample\cite{MorrisonBrowne1992}, then---as hinted in our opening paragraph---small-angle x-ray scatter\cite{GlatterKratky1982} (SAXS) will also be present.  This augments the previously mentioned coherent energy-flow vector at each point in space, with an ensemble (``SAXS fan'') of energy-flow vectors associated with diffusive energy transport.  How can this additional effect be taken into account, in modelling paraxial hard x-ray imaging of thin objects?

One answer is provided by a more general equation than the TIE, namely the the Fokker--Planck equation\cite{Risken1989}.  The latter equation, which can model flows in radiation and matter wave fields that have both a coherent and a diffusive component, may be obtained in a very general setting as a limit case of the still more general Boltzmann equation of classical statistical mechanics\cite{WangUhlenbeck1945}.  The Fokker--Planck equation is used in many fields: examples include Brownian motion \cite{WangUhlenbeck1945}, hydrodynamics\cite{Singh2015}, electron and photon transport in biological tissues \cite{OlbrantFrank2010, Osnabrugge2017} and homogeneous media\cite{Akcasu1997}, high-temperature atomic-level disorder in pre-melting surfaces \cite{Ferrando1992}, gaseous micro-flows \cite{Singh2016}, droplet nucleation \cite{Kuipers2010}, dilute polymer dynamics \cite{Singh2013}, hot plasmas \cite{Cooper1971}, biaxial fluids \cite{Kroger1997}, cold turbulent gas jets \cite{Naert1997} and quark gluon plasmas \cite{Selikhov1994}. Notable also is the use of the Fokker--Planck equation in the non-imaging context of x-ray kinematical and dynamical diffraction from imperfect crystals with stochastically distributed defects\cite{Davis1991,Davis1994}.  

The Fokker--Planck equation is also useful for x-ray imaging, since this is a setting where both coherent and diffusive energy flows are present.  This equation can simultaneously model four effects, resulting from the illumination of a thin object by normally incident coherent x-ray plane waves: 
\begin{itemize}
    \item Local absorption by the object (attenuation contrast, known since the 1895 discovery of x rays by R\"{o}ntgen\cite{Rontgen1896});
    \item Local lensing by the object which concentrates or rarefies energy density due to local focusing or defocusing (propagation-based Laplacian-type phase contrast\cite{Gureyev2006});
    \item  Local prism-like effects of the object which transversely shift optical energy (propagation-based differential phase contrast\cite{Gureyev2006});
    \item Local blurring with a position-dependent point spread function, associated with the SAXS fan emerging from each point at the exit of the object.
\end{itemize}
The first three effects are associated with the ``TIE part'' of the Fokker--Planck equation, with the fourth effect associated with the ``diffusion equation'' part.  Of particular interest is the use of an x-ray form of the Fokker--Planck equation for modelling forward problems in paraxial x-ray imaging in a simple manner, e.g.~by expressions valid in the near field (small propagation distance, Fresnel number\cite{SalehTeichBook} $N_{\textrm{F}}\gg 1$) that are of first-order accuracy with respect to the sample-to-detector propagation distance.  
A Kramers--Moyal\cite{Risken1989} extension will also be of use when a more detailed treatment of the SAXS, emanating from each point over the exit surface of the object, is required. We  anticipate that the resulting formalism will be of use in forward problems related to using x rays to image thin objects, as well as inverse problems such as phase/amplitude/diffusion-tensor recovery in both two and three dimensions (the latter constitutes the field of tensor tomography\cite{Gullberg1999,Malecki_2014,bayer2014reconstruction,schaff2015,liebi2015nanostructure,Wieczorek2016}).  

We close this introduction by outlining the remainder of the paper.  The x-ray Fokker--Planck model is derived in Sec.~2, in two complementary ways.   Section 2.1 considers the Fokker--Planck equation to be a fusing of (i) the transport-of-intensity equation for coherent energy transport in paraxial monochromatic x-ray beams, with (ii) the diffusion equation for diffusive paraxial energy transport associated with small-angle x-ray scattering within a sample.  The derivation of Sec.~2.2 is a microscopic treatment based on first principles.  Section 3 gives a generalised form of the Fokker--Planck equation, known as the Kramers--Moyal equation, as a means for treating the in-sample SAXS in a more precise manner.  Section 4 sketches some indicative means by which the Fokker--Planck and Kramers--Moyal equations may be used in the forward problem of modelling near-field intensity contrast in coherent x-ray imaging, together with the associated inverse problem of phase retrieval and SAXS determination. This section also considers some broader implications of the Fokker--Planck formalism.  We conclude with Sec.~5.      

\section{Fokker--Planck model for x-ray imaging}
\label{sec:FPM}

Here we give two complementary derivations of the x-ray Fokker--Planck equation.  The first is phenomenological, based on local conservation of energy in the presence of both coherent and diffusive paraxial x-ray energy transport.  The second is a microscopic first-principles analysis.  The former derivation is more general, while the latter is more physically illuminating.   

\subsection{X-ray Fokker--Planck equation: Transport-of-intensity equation with a diffusive term}
\label{sec:TIEAddition}

X-ray beams ``flow'' in the sense that they stream optical energy as they propagate.  As a descriptor of such energy flows, the Fokker--Planck equation may be viewed as a diffusive generalisation of coherent paraxial photon energy transport.  Adopting this perspective, below we consider the transport-of-intensity equation \cite{Teague1983} for coherent paraxial x-ray energy transport (Sec.~2.1.1), followed by the diffusion equation for diffusive paraxial x-ray energy transport (Sec.~2.1.2).  We then show that these two equations may be merged, in a manner that conserves energy both locally and globally, to give the Fokker--Planck equation for paraxial x-ray imaging (Sec.~2.1.3).

\subsubsection{Transport-of-intensity equation for coherent paraxial energy transport}

The transport-of-intensity equation \cite{Teague1983} describes the intensity evolution of a paraxial complex coherent scalar wave field as it propagates.  This continuity equation, arising from the parabolic equation of paraxial scalar wave optics, expresses local energy conservation.  It states that the negative divergence---which may loosely be spoken of as the ``convergence''---of the transverse Poynting vector (energy flow vector) is proportional to the longitudinal rate of change of intensity:\cite{Teague1983}
\begin{equation}
\frac{\partial}{\partial z}I(x,y,z)=-\frac{1}{k} \nabla_{\perp}\cdot[I(x,y,z)\nabla_{\perp}\phi(x,y,z)].
\label{eq:TIE}
\end{equation}
Here, $I$ is the intensity of the wave field, $\phi$ is its phase, $k$ is the wave-number and $\nabla_{\perp}$ is the gradient in $(x,y)$ planes perpendicular to the optic axis $z$.  Equation~(\ref{eq:TIE}) quantifies the fact that energy density (and hence intensity) increases with increasing propagation distance for locally-converging waves, and decreases with increasing propagation distance for locally-diverging waves.          

 As previously mentioned, the TIE is a continuity equation expressing local energy conservation under {\em coherent} x-ray energy transport. The conserved (Noether\cite{GoldsteinBook}) current is the transverse Poynting vector, which up to a multiplicative constant is\cite{GreenWolf1953,BerryOpticalCurrents,Paganin2006}
\begin{equation}
I(x,y,z) \, \nabla_{\perp}\phi(x,y,z)/k\equiv{\bf J}_{\perp}^{(1)}(x,y,z).
\end{equation}

For propagation {\em in vacuo} through a distance $\Delta \ge 0$ that is sufficiently small, a longitudinal finite-difference approximation to the TIE can be used to model the resulting propagation-based phase contrast\cite{Paganin2006}:
\begin{eqnarray}
\nonumber I(x,y,z=\Delta) \!\!\!\! &=& \!\!\!\!
I(x,y,z=0)-(\Delta/k)\{\nabla_{\perp}\cdot[I(x,y,z)\nabla_{\perp}\phi(x,y,z)]\}_{z=0} \\  &=& \!\!\!\! I(x,y,z=0)-(\Delta/k)\{   
I(x,y,z)\nabla_{\perp}^2 \phi(x,y,z)
+
\nabla_{\perp} I(x,y,z) \cdot \nabla_{\perp}\phi(x,y,z)
\}_{z=0}.
\label{eq:PropagationBasedPhaseContrast}
\end{eqnarray}

The underpinning physics is sketched in an imaging context in Fig.~\ref{fig:Samples}(a), whereby local specular refraction of x-ray radiation, by a thin object, leads to transverse re-distribution of optical energy upon propagation through a distance $\Delta$ that is not too large.  Here and henceforth, we use the term ``specular refraction'' in direct analogy to ``specular reflection'', with the former term referring to the refraction associated with a mirror-like  refracting surface (i.e.~smooth, lacking in the effects or roughness or spatial randomness; cf.~Latin {\em speculo}).  Rays passing through $A$ have a positive divergence, hence the intensity at $B$ is reduced compared to that which would have been registered if the rays passing through $A$ had all been parallel to the optic axis $z$.  Similarly, rays passing through $C$ have a negative divergence (positive ``convergence''), hence the intensity at $D$ is increased compared to that which would have been obtained had the rays exiting $C$ been parallel to the optic axis.  This phase contrast effect is increased with increased propagation distance $\Delta$, reduced x-ray energy (decreased wave-number $k$) or faster spatial variations in wave-field phase (larger $|\nabla_{\perp}\phi(x,y,z=0)|$).  We can also have effects such as a transverse deflection, e.g.~with the local maximum of intensity at $D$ being slightly shifted from the point $x=x_0$. Concentration or rarefaction of energy density (intensity) may be thought of as due to the lensing term proportional to $I(x,y,z)\nabla_{\perp}^2 \phi(x,y,z)$ in the bottom line of Eq.~(\ref{eq:PropagationBasedPhaseContrast}), while the transverse shifts in intensity maxima or minima may be viewed as due to the prism term $\nabla_{\perp}I(x,y,z)\cdot\nabla_{\perp} \phi(x,y,z)$.\cite{Gureyev2006}  

\subsubsection{Diffusion equation for diffusive paraxial energy transport}

If there were only diffusive energy transport on account of unresolved micro-structure in the sample (i.e.~SAXS), but no specular refraction, we could instead introduce a diffusion coefficient $D(x,y,z)$ to describe the effects at distance $z=\Delta$ downstream of an illuminated thin object in the plane $z=0$.  The fact that this coefficient depends on $x$ and $y$ reflects the fact that the SAXS fan will in general vary with position over the nominal exit surface $z=0$ of the thin sample.  Bearing all of this in mind, the intensity evolution of the beam may then be described by the diffusion equation:
\begin{equation}\label{eq:DiffusionEquation}
\left[\frac{\partial}{\partial z} I(x,y,z)\right]_{z=0}=\nabla_{\perp}^2[D(x,y,z=0) I(x,y,z=0)], \end{equation}
with the very important proviso that the above expression is always to be understood in its finite-difference form
\begin{equation}\label{eq:DiffusionEqnFiniteDifferenceForm}
I(x,y,z=\Delta)=I(x,y,z=0)+\Delta\left\{\nabla_{\perp}^2[D(x,y,z=0; \Delta) I(x,y,z=0)]\right\}_{z=0}, \quad \Delta \ge 0.\end{equation}
Note that a formal dependence on $\Delta$ has been added to the diffusion coefficient $D$, since SAXS-induced diffusive blur has a characteristic transverse width that scales with $\Delta$ rather than $\Delta^{1/2}$ (cf. Eqs~(e) and (f) in Fig.~3, the scaling relation in Eq.~(\ref{eq:InverseProblem3}), and the paragraph following Eq.~(\ref{eq:InverseProblem3})). 

The effect shown in one transverse dimension in Fig.~\ref{fig:Samples}(b) can be modelled using the above finite-difference form of the diffusion equation.  Here, an x-ray beamlet of width $\Delta x$ illuminates a slab containing SAXS-inducing unresolved micro-structure.  This width $\Delta x$ is assumed to be large compared to the characteristic length scale of the projected micro-structure, but small compared to the characteristic transverse width associated with the macroscopic projected phase and attenuation coefficients of the object.  The illuminated region $E$ of the slab, located at transverse coordinate $x=x_0$, results in a smooth SAXS fan of opening angle $\theta(x_0)$, which smears the propagated intensity over the plane $z=\Delta$.  The characteristic width of this blur, which may be considered as a broadening of the local rocking curve due to the sample\cite{Pagot2003,Wernick2003}, is
\begin{equation}
L(x_0;\Delta)\approx \theta(x_0)\Delta\approx \sqrt{D(x_0,z=0;\Delta)\,\Delta}.   
\end{equation}
Thus, local diffuse scattering by a thin object leads to a transverse smearing of intensity upon propagation.  This blur increases with increased propagation distance $\Delta$.  At the level at which diffusion is here being considered, the opening angle $\theta(x_0)$ of the SAXS fan is taken to completely characterise its diffusive effects at each location $x_0$; a more sophisticated treatment which replaces the single number $\theta(x_0)$ (or its associated diffusion coefficient) with a hierarchy of tensors, will be given in Sec.~3.

Like the TIE, the diffusion equation may be viewed as a continuity equation expressing local energy conservation, albeit under {\em diffusive} x-ray energy transport.  Here, the conserved current is given by Fick's first law as\cite{Risken1989,CrankBook}
\begin{equation}
-\nabla_{\perp}[D(x,y,z=0;\Delta) I(x,y,z=0)]\equiv{\bf J}_{\perp}^{(2)}(x,y,z=0).
\end{equation}

\subsubsection{Fokker--Planck equation for combined coherent and diffusive paraxial energy transport}

If both specular refraction and local-SAXS effects are present simultaneously, we can add the TIE and diffusion equations together, in an energy-preserving manner.  This gives the continuity equation:
\begin{eqnarray}\label{eq:FokkerPlanckEquationAbstractForm}
  \left[\frac{\partial}{\partial z} I(x,y,z)\right]_{z=0} &=& -\nabla_{\perp}\cdot{\bf J}_{\perp}(x,y,z=0), \\ \nonumber \quad{\bf J}_{\perp}(x,y,z=0) &=& [1-F(x,y)] \, {\bf J}_{\perp}^{(1)}(x,y,z=0)+F(x,y) \, {\bf J}_{\perp}^{(2)}(x,y,z=0), \quad 0\le F(x,y)\le 1.
\end{eqnarray}
Here, $F(x,y)$ is the fraction of the optical energy converted to SAXS\cite{Strobl2014} when illuminating a thin sample in the plane $z=0$ at the point $(x,y)$.  If $F(x,y)\ll 1$ (as is often the case for both soft\cite{MorrisonBrowne1992} and hard\cite{SuzukiUchida1995} x-rays) one can replace $1-F(x,y)$ in the above expression by $1$.  The above convection--diffusion equation (forward Kolmogorov equation\cite{Risken1989}) is the Fokker--Planck equation in two transverse dimensions.  It models {\em both} effects in Fig.~\ref{fig:Samples}, i.e.~it models the effects of both local sample-induced specular refraction and local sample-induced SAXS.  

If the expressions for the coherent current ${\bf J}_{\perp}^{(1)}$ and diffusive current ${\bf J}_{\perp}^{(2)}$ are written explicitly, and the approximations made that (i) $F(x,y)\ll 1$ and (ii) $F(x,y)$ is slowly varying with $x$ and $y$, we obtain the 2+1-dimensional Fokker--Planck equation:
\begin{eqnarray}\label{eq:FPEin2D}
\frac{\partial}{\partial z}I(x,y,z)=-\frac{1}{k}\nabla_{\perp}\cdot[I(x,y,z)\nabla_{\perp}\phi(x,y,z)]+F(x,y)\nabla_{\perp}^2 \left[D(x,y,z=0;\Delta)I(x,y,z)\right].
\end{eqnarray}
In one transverse dimension, this becomes the one-dimensional Fokker--Planck equation:
\begin{eqnarray}\label{eq:FokkerPlanckXray}
\frac{\partial}{\partial z}I(x,y,z) = -\frac{1}{k}\frac{\partial}{\partial x} \left[I(x,z)\frac{\partial}{\partial x}\phi(x,z)\right]+F(x)\frac{\partial^2}{\partial x^2} \left[D(x,z=0;\Delta)I(x,z)\right].
\end{eqnarray}

The corresponding finite-difference forms, which are to be preferred for reasons that have already been outlined, are respectively given by the following pair of equations:
\begin{eqnarray}\label{eq:FPEin2D_finite_difference}
I(x,y,z=\Delta)=I(x,y,z=0)-(\Delta/k)\nabla_{\perp}\cdot[I(x,y,z)\nabla_{\perp}\phi(x,y,z)]_{z=0}+F(x,y)\Delta\nabla_{\perp}^2 \left[D(x,y,z=0;\Delta)I(x,y,z)\right]_{z=0},
\end{eqnarray}
\begin{eqnarray}\label{eq:FokkerPlanckXray_finite_difference}
I(x,z=\Delta)=I(x,z=0)-\frac{\Delta}{k}\frac{\partial}{\partial x} \left[I(x,z)\frac{\partial}{\partial x}\phi(x,z)\right]_{z=0}+F(x)\Delta\frac{\partial^2}{\partial x^2} \left[D(x,z=0;\Delta)I(x,z)\right]_{z=0}.
\end{eqnarray}

We close this section by noting that, if $D$ may be considered to be sufficiently slowly varying with respect to transverse coordinates that $D$ commutes with the transverse Laplacian, we may instead work with the simpler finite-difference expressions:
\begin{eqnarray}\label{eq:FPEin2D_finite_difference2}
I(x,y,z=\Delta)=I(x,y,z=0)-(\Delta/k)\nabla_{\perp}\cdot[I(x,y,z)\nabla_{\perp}\phi(x,y,z)]_{z=0}+F(x,y)D(x,y,z=0;\Delta)\Delta\nabla_{\perp}^2 I(x,y,z=0),
\end{eqnarray}
\begin{eqnarray}\label{eq:FokkerPlanckXray_finite_difference2}
I(x,z=\Delta)=I(x,z=0)-\frac{\Delta}{k}\frac{\partial}{\partial x} \left[I(x,z)\frac{\partial}{\partial x}\phi(x,z)\right]_{z=0}+F(x)D(x,z=0;\Delta)\Delta\frac{\partial^2}{\partial x^2} I(x,z=0).
\end{eqnarray}
In the above two finite-difference Fokker--Planck equations, $FD$ plays the role of an effective diffusion coefficient 
\begin{eqnarray}\label{eq:EffectiveDiffusionCoefficient}
D_{\textrm{eff}}(x,y,\Delta)=F(x,y) D(x,y,z=0;\Delta)
\end{eqnarray}
that also accounts for the fraction of the incident radiation converted to SAXS (cf.~Eq.~(\ref{eq:DiffusionCoeff2}), together with Eq.~(g) in Fig.~\ref{fig:RulesOfThumb}).

\subsection{X-ray Fokker--Planck equation: Derived from first principles}
\label{sec:FullDerivation}

Here we derive the x-ray Fokker--Planck equation from first principles.  We consider two length scales of sample-induced wave-field variation.  The larger length scale corresponds to detector-resolved phase and amplitude variations, and the smaller length scale to unresolved micro-structure resulting in SAXS.  This derivation clarifies the physical mechanisms underpinning both coherent and diffusive x-ray energy transport, in a paraxial x-ray imaging context.

We work in one transverse dimension for simplicity, ignore polarisation and assume monochromatic x rays.  Let $\psi(x,z=0)$ be the spatial part of the complex scalar amplitude at the exit surface of a sample.  Subsequent Fresnel diffraction gives\cite{Paganin2006} 
\begin{equation}\label{eq:Appendix01}
    \psi(x,z=\Delta)= \int P_{\Delta}(x')  \psi(x-x',z=0) \, dx',
\end{equation}
where $P_{\Delta}(x)$ is the Fresnel propagator corresponding to free-space diffraction through a distance $\Delta\ge 0$, and paraxial radiation is assumed throughout.  The squared modulus of the above convolution integral gives the propagated intensity as the following linear combination of contributions from every pair of points $x',x''$ in the unpropagated wave field\cite{Nesterets2008}:
\begin{equation}\label{eq:Appendix02}
    I(x,z=\Delta)\equiv|\psi(x,z=\Delta)|^2=\iint P_{\Delta}(x')P_{\Delta}^*(x'')\psi(x-x',z=0)\psi^*(x-x'',z=0) \, dx'dx''.
\end{equation}

The sample is assumed to be thin, normal to the incident plane-wave illumination, and with nominal exit surface $z=0$. Many models exist for incorporating unresolved micro-structure into this sample\cite{BeckmannSpizzichinoBook,VoronovichBook}.  Follow Voronovich \cite{VoronovichBook}, Nesterets\cite{Nesterets2008} and Yashiro et al.\cite{Yashiro2010,Yashiro2011} in decomposing the phase $\phi(x,z=0)$ of $\psi(x,z=0)$ as 
\begin{equation}
\phi(x,z=0)=\phi_{\textrm{s}}(x)+\phi_{\textrm{f}}(x), 
\end{equation}
where (i) $\phi_{\textrm{s}}(x)$ is a phase component that varies slowly with respect to the detector pixel size, and (ii) $\phi_{\textrm{f}}(x)$ is a fast-varying random phase due to the sample micro-structure, which fluctuates many times over one detector pixel and is thus unresolved (see also Gureyev et al.\cite{Gureyev2006}). Further assume a slowly-varying intensity $I_s(x)$, together with the projection approximation.  Hence
\begin{equation}
\psi(x,z=0)=\sqrt{I_s(x)}\exp\{i[\phi_{\textrm{s}}(x)+\phi_{\textrm{f}}(x)]\}, 
\end{equation}
and so Eq.~(\ref{eq:Appendix02}) becomes\cite{Nesterets2008}:
\begin{equation}\label{eq:Appendix03}
    I(x,z=\Delta)=\iint P_{\Delta}(x')P_{\Delta}^*(x'')\sqrt{I_s(x-x')I_s(x-x'')} \exp\{i[\phi_{\textrm{s}}(x-x')-\phi_{\textrm{s}}(x-x'')]\} \exp\{i[\phi_{\textrm{f}}(x-x')-\phi_{\textrm{f}}(x-x'')]\} \, dx'dx''.
\end{equation}
Average over an ensemble of realisations of the unresolved fast phase-maps $\phi_{\textrm{f}}(x)$ \cite{Pedersen1976,Nesterets2008, GoodmanSpeckleBook,VartanyantsRobinson2003}.  Denoting this average by an over-line,  
\begin{equation}\label{eq:Appendix04}
    \overline{I}(x,z=\Delta)=\iint P_{\Delta}(x')P_{\Delta}^*(x'')\sqrt{I_s(x-x')I_s(x-x'')} \exp\{i[\phi_{\textrm{s}}(x-x')-\phi_{\textrm{s}}(x-x'')]\} \overline{\exp\{i[\phi_{\textrm{f}}(x-x')-\phi_{\textrm{f}}(x-x'')]\}} \, dx'dx''.
\end{equation}
Assuming the fast phase maps $\phi_{\textrm{f}}$ to be spatially statistically stationary {\em over transverse displacements that are small compared to the characteristic length scale over which $\phi_{\textrm{s}}$ fluctuates appreciably}, the over-lined expression in Eq.~(\ref{eq:Appendix04}) depends only on coordinate differences $x'-x''$.  Assuming $\phi_{\textrm{f}}(x-x')$ and $\phi_{\textrm{f}}(x-x'')$ to be Gaussian variables, this correlation function is\cite{GoodmanStatisticalOpticsBook,VartanyantsRobinson2003,GoodmanSpeckleBook,Nesterets2008,Yashiro2010,Yashiro2011}
\begin{equation}\label{eq:Appendix05}
\overline{\exp\{i[\phi_{\textrm{f}}(x-x')-\phi_{\textrm{f}}(x-x'')]\}} =\exp\{-\sigma_{\phi_{\textrm{f}}}^2(x)[1-\gamma(|x'-x''|;x)]\},
\end{equation}
where $\sigma_{\phi_{\textrm{f}}}^2(x)$ is the $x$-dependent variance of the ensemble of fast phase maps $\{\phi_{\textrm{f}}(x)\}$, and 
\begin{equation}\label{eq:Appendix06}
    \gamma(\Delta x; x)= \frac{\overline{\phi_{\textrm{f}}(x)\phi_{\textrm{f}}(x+\Delta x)}}{\sigma_{\phi_{\textrm{f}}}^2(x)}
\end{equation}
is the normalised fast-phase auto-correlation function\cite{Lynch2011} as a function of coordinate separation $\Delta x$ and transverse coordinate $x$. 

\begin{figure}
\centering
\includegraphics[width=470pt]{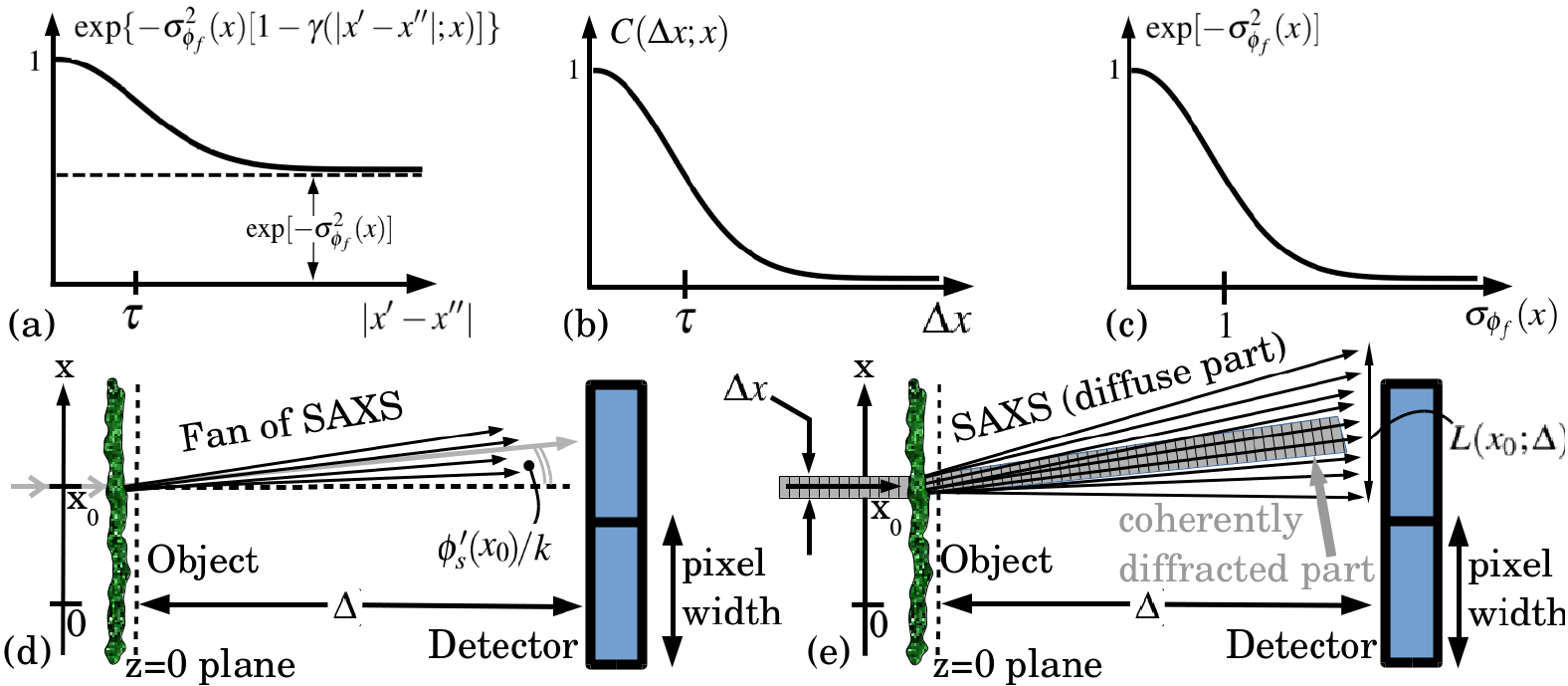}
\caption{(a) Auto-correlation of $\exp[i{\phi_{\textrm{f}}(x)}]$ in Eq.~(\ref{eq:Appendix07}), with $x$ considered fixed; (b) Re-normalised fast-phase correlation function in Eq.~(\ref{eq:Appendix08}), for a given fixed $x$; (c) Decoherence factor $\exp[-\sigma_{\phi_{\textrm{f}}}^2(x)]$ appearing in Eq.~(\ref{eq:Appendix09}) and onwards, for a given fixed $x$; (d) Physical interpretation of Eq.~(\ref{eq:Appendix11}); (e) Specular and diffuse scattering for a single x-ray beamlet of width $\Delta x$.}  
\label{fig:appendix}
\end{figure}

The quantity $\sigma_{\phi_{\textrm{f}}}(x)$ is called the ``phase depth''.  If this is not significantly greater than unity, the right side of Eq.~(\ref{eq:Appendix05}) does not decay to zero as $|x'-x''|\rightarrow\infty$.  Rather, this correlation function asymptotes to $\exp[-\sigma^2_{\phi_{\textrm{f}}}(x)]$: see Fig.~\ref{fig:appendix}(a).  The physical reason for this decay to a non-zero constant is the phase depth not being sufficiently high for $\exp(i\phi_{\textrm{f}})$ at two different locations to cross-correlate to zero, irrespective of how widely separated these locations may be (cf.~Fig.~3 in Prade et al.\cite{prade2016short}).      

To proceed further, follow Goodman\cite{GoodmanSpeckleBook} in separating out the asymptote term from $\overline{\exp\{i[\phi_{\textrm{f}}(x-x')-\phi_{\textrm{f}}(x-x'')]\}}$ in Eq.~(\ref{eq:Appendix05}), and re-normalising the part of the correlation that does decay to zero.  Thus:
\begin{equation}\label{eq:Appendix07}
\overline{\exp\{i[\phi_{\textrm{f}}(x-x')-\phi_{\textrm{f}}(x-x'')]\}}= \exp[-\sigma^2_{\phi_{\textrm{f}}}(x)]+\{1-\exp[-\sigma^2_{\phi_{\textrm{f}}}(x)]\} \, C(x'-x'';x),   
\end{equation}
where $C(x'-x'';x)$ is a re-normalised correlation function
\begin{equation}\label{eq:Appendix08}
C(x'-x'';x)=\frac{\exp\{-\sigma_{\phi_{\textrm{f}}}^2(x)[1-\gamma(|x'-x''|;x)]\}-\exp[-\sigma^2_{\phi_{\textrm{f}}}(x)]}{1-\exp[-\sigma^2_{\phi_{\textrm{f}}}(x)]}
\end{equation}
that does decay to zero for coordinate separations $\Delta x \equiv|x'-x''|\gg \tau$, and  $\tau$ is the transverse correlation length (see Fig.~\ref{fig:appendix}(b)).

Substituting Eq.~(\ref{eq:Appendix07}) into Eq.~(\ref{eq:Appendix04}) gives\cite{Nesterets2008}
\begin{eqnarray}\label{eq:Appendix09}
\overline{I}(x,z=\Delta) \!\!\!\!\!\! &=& \!\!\!\!\!\! \exp[-\sigma^2_{\phi_{\textrm{f}}}(x)]|\mathcal{D}_{\Delta}^{\textrm{(F)}}\{\sqrt{I_s(x)}\exp[i\phi_{\textrm{s}}(x)]\}|^2+\mathcal{B}(x,z=\Delta), \quad\quad\quad\quad\quad\quad\quad\quad\quad\quad\quad\quad  \\ \label{eq:Appendix10}
\mathcal{B}(x,z=\Delta) \!\!\!\!\!\! &=& \!\!\!\!\!\! \nonumber\{1-\exp[-\sigma^2_{\phi_{\textrm{f}}}(x)]\} \\ &\times& \!\!\!\!\!\! \iint P_{\Delta}(x')P_{\Delta}^*(x'')\sqrt{I_s(x-x')I_s(x-x'')} \, \exp\{i[\phi_{\textrm{s}}(x-x')-\phi_{\textrm{s}}(x-x'')]\} \, C(x'-x'';x)dx'dx'', 
\end{eqnarray}
where $\mathcal{D}_{\Delta}^{\textrm{(F)}}$ is the Fresnel diffraction operator\cite{Paganin2006} acting on the smooth component of the complex amplitude, and $\mathcal{B}(x,z=\Delta)$ represents small-angle x-ray scattering.  Thus a physical consequence of the lack of decay to zero in Eq.~(\ref{eq:Appendix05}), is that the diffraction pattern in Eq.~(\ref{eq:Appendix09}) contains two components\cite{Beckmann1967,Sinha1988,VartanyantsRobinson2003,GoodmanSpeckleBook,Nesterets2008}: 
\begin{itemize}
    \item a ``specular diffraction'' component that undergoes Fresnel diffraction as if there were no micro-structure present, but with intensity damped by the decoherence factor\cite{Nesterets2008} shown in Fig.~\ref{fig:appendix}(c);
    \item a diffuse component $\mathcal{B}(x,z=\Delta)$ associated with SAXS.
\end{itemize}
 This splitting of x-ray energy flows---whereby an incident coherent x-ray energy flow upstream of the object bifurcates into superposed coherent and diffusive energy flows downstream of the object---corresponds directly to the factors of $F$ and $1-F$ introduced ``by hand'' in the previous sub-section (see Eq.~(\ref{eq:FokkerPlanckEquationAbstractForm})), if we take 
\begin{equation}\label{eq:RelationBetweenPhaseDepthandF}
    F(x)=1-\exp[-\sigma^2_{\phi_{\textrm{f}}}(x)]\xrightarrow[\sigma^2_{\phi_{\textrm{f}}}(x)\ll 1]{\textrm{weak SAXS}}1-[1-\sigma^2_{\phi_{\textrm{f}}}(x)]=\sigma^2_{\phi_{\textrm{f}}}(x).
\end{equation}
Note that a similar bifurcation of energy flows occurs, for similar reasons, in statistical dynamical x-ray diffraction theory\cite{Kato1,Kato2}.  Note also that a non-zero specular-diffraction component (which will typically be the dominant component in the present context) indicates the SAXS-induced speckle to ``partially developed'' \cite{Pedersen1974}; this may be contrasted with the case of ``fully developed'' SAXS speckle, which is not relevant to our context, in which the specular component would be fully extinguished.

Assume $I_s(x)$ to be sufficiently slowly varying that the square-root term in Eq.~(\ref{eq:Appendix10}) may be replaced by $I_s(x)$. For the complex exponent in Eq.~(\ref{eq:Appendix10}), make the more sensitive approximation:
\begin{equation}
\phi_{\textrm{s}}(x-x')-\phi_{\textrm{s}}(x-x'')\approx(x''-x')\phi_{\textrm{s}}'(x), 
\end{equation}
where $\phi_{\textrm{s}}'(x)$ denotes the derivative of $\phi_{\textrm{s}}(x)$ with respect to $x$; the above equation makes implicit use of the fact that $\phi_{\textrm{f}}$ varies much more quickly than $\phi_{\textrm{s}}$.  Then write $P_{\Delta}(x')$ in Eq.~(\ref{eq:Appendix10}) as a Fourier integral using the notation and convention specified by
\begin{equation}
P_{\Delta}(x')=(2\pi)^{-1/2}\int\breve{P}_{\Delta}(k_x)\exp(ik_xx')dk_x,    
\end{equation}
where $k_x$ is the Fourier coordinate corresponding to $x$.  Also write $P^*_{\Delta}$ and $C$ in Eq.~(\ref{eq:Appendix10}) as Fourier integrals, with $C$ being Fourier transformed with respect to its first argument.  With all of these steps, together with the fact that $|\breve{P}_{\Delta}|=1$ for the Fresnel propagator\cite{Paganin2006}, one obtains (cf.~similar equations in related contexts\cite{Szoke2001,Borowski2001,Nesterets2008,VartanyantsRobinson2003}, especially those due to Nesterets\cite{Nesterets2008}):
\begin{equation}\label{eq:Appendix11}
  \mathcal{B}(x,z=\Delta)=\sqrt{2\pi} \, \{1-\exp[-\sigma^2_{\phi_{\textrm{f}}}(x)]\} I_s(x) \int dk_x \, \breve{C}(\phi_{\textrm{s}}'(x)-k_x;x).  
\end{equation}

Physically, the above equation states that the SAXS contribution, at a given transverse location $x$, integrates over a fan of rays making angles $[\phi_{\textrm{s}}'(x)-k_x]/k$ with respect to the optic axis.  This fan of rays is due to scattering from unresolved micro-structure associated with the ensemble of fast phase maps $\{\phi_{\textrm{f}}(x)\}$.  The whole fan of SAXS rays is rotated by the angle $\phi_{\textrm{s}}'(x)/k$, due to the specular refraction. All SAXS-scattered x rays, due to illumination of the point at $x=x_0$, are integrated over a single pixel. See Fig.~\ref{fig:appendix}(d).  The angular spread of the SAXS fan will typically be on the order of a few degrees (50 milli-radians) for hard x rays\cite{HeBook2009}, while the fan rotation angle $\phi_{\textrm{s}}'(x)/k$ will typically be on the order of micro-radians\cite{FitzgeraldPhysicsToday2000}.  The function $\breve{C}$, being the Fourier transform of a correlation function\cite{Strobl2014}, is proportional to the power spectrum of the ensemble of fast-phase fields $\exp\{i[\phi_{\textrm{f}}(x)]\}$ (Wiener--Khinchin theorem\cite{BornWolf}); this power spectrum (local differential scattering cross section\cite{HardingSchreiber1999,Lynch2011,prade2016short}; cf.~Eq.~(\ref{eq:KramersMoyal2}) below) gives the weights for the diffracted ``rays'' in the SAXS fan.  The fraction $F(x)=1-\exp[-\sigma^2_{\phi_{\textrm{f}}}(x)]$ of the beam converted to SAXS is complementary to the fraction $1-F(x)=\exp[-\sigma^2_{\phi_{\textrm{f}}}(x)]$ channelled into the specular Fresnel diffraction (cf.~Eq.~(\ref{eq:RelationBetweenPhaseDepthandF})).

If the object-to-detector distance $\Delta$ is large enough that the SAXS-induced fan of rays in  Fig.~\ref{fig:appendix}(d) spreads over more than one pixel, we have the scenario shown in  Fig.~\ref{fig:appendix}(e).  Equation~(\ref{eq:Appendix11}) generalises to:
\begin{equation}\label{eq:Appendix12}
  \mathcal{B}(x=x_0,z=\Delta)=\sqrt{2\pi} \, \{1-\exp[-\sigma^2_{\phi_{\textrm{f}}}(x)]\} I_s(x) \star  d(x;x_0),  \end{equation}
where $\star$ denotes convolution over the $x$ variable, and $d(x;x_0)$ is a diffusive blur kernel given by 
\begin{equation}
d(x;x_0)=\breve{C}(k_x;x_0)_{k_x=\phi_{\textrm{s}}'(x)-kx/\Delta}\approx \breve{C}(k_x;x_0)_{k_x=-kx/\Delta}.
\end{equation}
The function $d$ is simply a transverse re-scaling of  $\breve{C}$, from a function of Fourier-space coordinates $k_x$ (i.e.~angular coordinates $k_x/k$, in the paraxial approximation) to a function of the transverse coordinates $x$ over the detector plane.  Stated differently,  $d(x;x_0)$ is the intensity distribution of SAXS, as a function of position $x$ on the detector, due to x rays incident at position $x_0$.  
If the object-to detector distance $\Delta$ is sufficiently small, two additional approximations can be made.  

(i) Small $\Delta$ implies $d(x;x_0)$ to be sufficiently narrow for its Fourier transform $\breve{d}(k_x;x_0)$ with respect to $x$ to be sufficiently broad that it can be represented by a second-order Taylor expansion in  $k_x$.  The term in the Taylor series for $\breve{d}(k_x;x_0)$ that is linear in $k_x$ will vanish since $C(\Delta x;x_0)$ is approximately even in $\Delta x$ (cf.~Eq.~(\ref{eq:SAXS-fan-centroid-is-at-the-origin})).  Given all of the above, the convolution operator  $\sqrt{2\pi} \,  d(x;x_0)\star$ in Eq.~(\ref{eq:Appendix12}) is well approximated by
\begin{equation}\label{eq:BlurOperatorSecondOrder}
\sqrt{2\pi} \, d(x;x_0)\star\approx Q+L^2(x_0;\Delta)(d^2/dx^2)  ,
\end{equation}
where $L(x_0;\Delta)$ is an $x_0$-dependent blurring width (this blurring being due to SAXS), and $Q$ is a positive constant.  Invoke conservation of energy, which implies that the total scattered intensity---due to an incident x-ray pencil beam of intensity $I_s(x)$ and with small thickness $\Delta x$---will have the same integrated intensity after scattering from the object; see Fig.~\ref{fig:appendix}(e).  This shows that $Q=1$, so that $\sqrt{2\pi} \, d(x;x_0)\star\approx 1+L^2(x_0;\Delta)(d^2/dx^2)$ (cf.~Gureyev et al.\cite{DeblurByDefocus}).  Hence we may approximate Eq.~(\ref{eq:Appendix12}) as:
\begin{equation}\label{eq:Appendix13}
  \mathcal{B}(x=x_0,z=\Delta)=\{1-\exp[-\sigma^2_{\phi_{\textrm{f}}}(x)]\} \left[1+L^2(x_0;\Delta)\frac{d^2}{dx^2}\right] I_s(x).  \end{equation}

Before proceeding, let us back-track a little, and point out that in Eq.~(\ref{eq:Appendix12}) we consider $I_s(x) \star  d(x;x_0)$ to be equivalent to $\int I_s(x_0) d(x-x_0;x_0) \, dx_0$ .  Strictly speaking, this linear integral transform is not a convolution integral since the blurring kernel $d$ depends on $x_0$.  We use the convolution notation for clarity, however, to denote such ``smearing with a position dependent point spread function that results from the position-dependent SAXS''.  See also Sec.~3, whose analysis considers the same linear integral transform using a more conventional and explicit notation.

(ii) A second consequence of small $\Delta$ is that the TIE approximation\cite{Teague1983} may be applied to the first term on the right of Eq.~(\ref{eq:Appendix09}):
\begin{eqnarray}\label{eq:Appendix14}
|\mathcal{D}_{\Delta}^{\textrm{(F)}}\{\sqrt{I_s(x)}\exp[i\phi_{\textrm{s}}(x)]\}|^2\approx I_s(x)-\frac{\Delta}{k}\frac{d}{d x}\left[ I_s(x) \frac{d \phi_{\textrm{s}}(x)}{d x}  \right]. \end{eqnarray}
If Eqs~(\ref{eq:Appendix13}) and (\ref{eq:Appendix14}) are substituted into Eq.~(\ref{eq:Appendix09}), we obtain
\begin{eqnarray}\label{eq:Appendix15}
\frac{\overline{I}(x,z=\Delta)-I_s(x)}{\Delta}=\exp[-\sigma^2_{\phi_{\textrm{f}}}(x)] \left(\frac{-1}{k}\right)\frac{d}{d x}\left[ I_s(x) \frac{d\phi_{\textrm{s}}(x)}{d x}  \right]+\frac{1}{\Delta}\{1-\exp[-\sigma^2_{\phi_{\textrm{f}}}(x)]\}L^2(x;\Delta)\frac{d^2 I_s(x)}{d x^2}.
\end{eqnarray}
Assume only a small fraction of the incident x rays to be converted to SAXS, so that  $\sigma^2_{\phi_{\textrm{f}}}(x)\ll 1$.  Thus:
\begin{eqnarray}\label{eq:Appendix16}
\frac{\overline{I}(x,z=\Delta)-I_s(x)}{\Delta}\approx 
-\frac{1}{k}\frac{d}{d x}\left[ I_s(x) \frac{d\phi_{\textrm{s}}(x)}{d x}  \right]+\frac{L^2(x)}{\Delta} \sigma^2_{\phi_{\textrm{f}}}(x) \frac{d^2 I_s(x)}{d x^2}.
\end{eqnarray}
Upon comparing the final term of Eq.~(\ref{eq:Appendix16}) with the final term in Eq.~(\ref{eq:FPEin2D_finite_difference2}), and then noting from the far right side of Eq.~(\ref{eq:RelationBetweenPhaseDepthandF}) that $\sigma^2_{\phi_{\textrm{f}}}(x)\approx F(x)$, we obtain:
\begin{equation}
D(x,z=0;\Delta)=L^2(x;\Delta)/\Delta.
\label{eq:DiffusionCoeff1}
\end{equation}
The corresponding position-dependent effective diffusion coefficient $D_{\textrm{eff}}(x;\Delta)$ obeys Eq.~(\ref{eq:EffectiveDiffusionCoefficient}), so that
\begin{equation}
D_{\textrm{eff}}(x;\Delta)=\sigma^2_{\phi_{\textrm{f}}}(x) D(x,z=0;\Delta)=\sigma^2_{\phi_{\textrm{f}}}(x) L^2(x;\Delta)/\Delta. 
\label{eq:DiffusionCoeff2}
\end{equation}
Equation~(\ref{eq:Appendix16}) then becomes the 1D form of the finite-difference x-ray Fokker--Planck equation in Eq.~(\ref{eq:FokkerPlanckXray_finite_difference}), provided the diffusion coefficient varies sufficiently slowly with $x$ that it commutes with the transverse Laplacian.   Note also that a Kramers--Moyal generalisation of the Fokker--Planck equation would have resulted if all terms had been included in the expansion for $\sqrt{2\pi} \, d(x;x_0)\star$ (see Sec.~3).  We again emphasise that Eq.~(\ref{eq:Appendix16}) is to be considered with $\Delta$ being {\em small but finite and fixed}.  Thus, the simplicity of using diffusion concepts to describe SAXS, must be balanced against the fact that free-space propagation blur-width $L$ scales as $\Delta$ rather than $\sqrt{\Delta}$; formally, this is accounted for by having a $\Delta$-dependent diffusion coefficient for the x-ray Fokker--Planck equation, {\em which is only ever used in its finite-difference form for small nonzero $\Delta$}. 

\begin{figure}[!h]
\includegraphics[width=350pt]{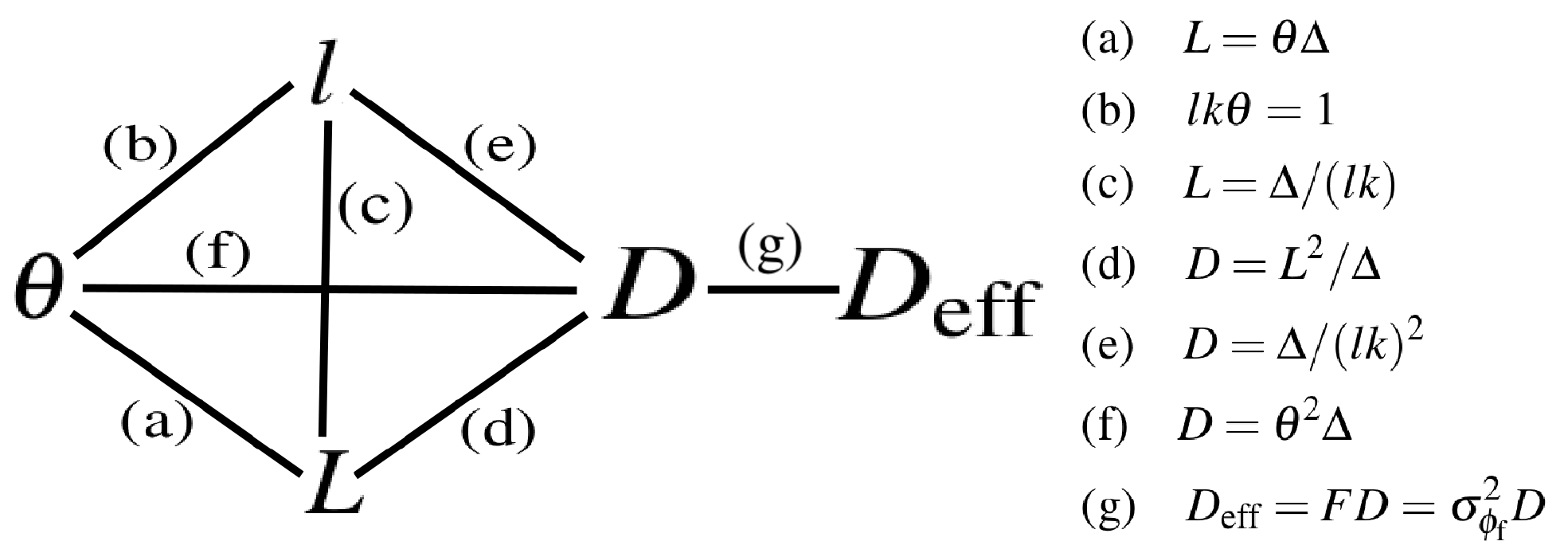}
\centering
\caption{Relations between several key physical quantities used in this paper.}
\label{fig:RulesOfThumb}
\end{figure}

We close this section with Fig.~\ref{fig:RulesOfThumb}.  This indicates the relation between various key quantities: the divergence angle $\theta$ of the SAXS fan, the characteristic transverse length scale $l$ of the spatially-rapid wave-front fluctuations induced by unresolved micro-structure at the sample exit surface, the phase depth $\sigma_{\phi_{\textrm{f}}}$ of the associated spatially-rapid phase fluctuations, the diffusion coefficient $D$ and the effective diffusion coefficient $D_{\textrm{eff}}$ in the Fokker--Planck equation, and the smear width $L$ associated with the SAXS fan. We make the following remarks on each equation in Fig.~\ref{fig:RulesOfThumb}.

\bigskip

(a) The link between $L$ and $\theta$ follows from elementary geometry; 

(b) The link between $l$ and $\theta$ follows from the optical uncertainty principle \cite{Bracewellbook}; 

(c) The link between $l$ and $L$ follows upon eliminating $\theta$ from relations (a) and (b); 

(d) The link between $D$ and $L$ follows from Eqs~(\ref{eq:DiffusionCoeff1}) and (\ref{eq:DiffusionCoeff2}); 

(e) The link between $D$ and $l$ follows upon eliminating $L$ from relations (c) and (d); 

(f) The link between $D$ and $\theta$ follows upon eliminating $L$ from relations (a) and (d);

(g) The link between $D$ and $D_{\textrm{eff}}$ follows from Eqs~(\ref{eq:EffectiveDiffusionCoefficient}) and (\ref{eq:DiffusionCoeff2}); see also Eq.~(\ref{eq:RelationBetweenPhaseDepthandF}).


\bigskip

\noindent All physical dependencies, for the quantities in Fig.~\ref{fig:RulesOfThumb}, are intuitively reasonable: (a) the blur width $L$ is directly proportional to the angular spread $\theta$ of the SAXS fan; the blur width $L$ is directly proportional to the sample-to-detector distance $\Delta$; (b) the angular width $\theta$ of the SAXS fan is inversely proportional to the transverse length scale $l$ of the projected unresolved micro-structure, etc.

\section{Kramers--Moyal extension of x-ray Fokker--Planck equation}

The preceding analysis gives only a crude treatment of the angular distribution of the local SAXS fan, which is described merely via its angular spread $\theta(x,y)$ at each point over the exit surface $z=0$ of a thin sample.  Here, we show that a less-crude treatment leads to the Kramers--Moyal generalisation of the Fokker--Planck equation, in a form suitable for x-ray near-field imaging of thin objects.

Recall Eq.~(\ref{eq:RelationBetweenPhaseDepthandF}), which links the local phase depth associated with the projected micro-structure of a thin object, to the associated fraction $F(x,y)$ of the x-ray beam which is channelled into SAXS.  The two-transverse-dimensional forms of Eqs~(\ref{eq:Appendix09}),  (\ref{eq:Appendix12}) and (\ref{eq:Appendix14}), upon writing the linear integral transform in Eq.~(\ref{eq:Appendix12}) explicitly and setting $\sqrt{2\pi}d\equiv K$, then become:
\begin{eqnarray} \label{eq:KramersMoyal1}
    \overline{I}(x,y,z=\Delta) \!\!\!\!\!\! &=& \!\!\!\!\!\! \left[1-F(x,y)\right] \left\{  I_s(x,y,z=0)-(\Delta/k)\nabla_{\perp}\cdot[I_s(x,y,z)\nabla_{\perp}\phi_{\textrm{s}}(x,y,z)]_{z=0} \right\} + \mathcal{B}(x,y,z=\Delta), \\ \label{eq:KramersMoyal1b} \mathcal{B}(x,y,z=\Delta) \!\!\!\!\!\! &=& \!\!\!\!\!\! F(x,y) \! \iint I_s(x',y',z=0) \, K (x,y,x',y',z=\Delta) \, dx'dy'.
\end{eqnarray}
Again, the first term on the right side of Eq.~(\ref{eq:KramersMoyal1}) is associated with the coherently-scattered component of the x-ray field downstream of the thin object in the plane $z=0$, while the second term $\mathcal{B}$ is associated with the diffusely scattered component.    

Recall the usual definition of a differential scattering cross section $[d\sigma/d\Omega](\theta_x,\theta_y)$, where $d\Omega$ is an infinitesimal element of solid angle and $\theta_x,\theta_y$ describe angular deflections (scattering angles) in the $x,y$ transverse directions, respectively\cite{MartinShaw1997}. Hence conclude that $K=\sqrt{2\pi}d$ in Eq.~(\ref{eq:KramersMoyal1b}) is proportional to the local differential scattering cross section\cite{HardingSchreiber1999} associated with unresolved micro-structure in the sample, namely a differential scattering cross section $[d\sigma/d\Omega](x',y';\theta_x,\theta_y)$ that depends upon an illuminating beamlet's centre $(x',y')$, as well as upon the scattering angles $\theta_x,\theta_y$.  Thus we have the proportionality
\begin{equation}\label{eq:KramersMoyal2}
    K (x,y,x',y',z=\Delta)\propto \frac{d\sigma}{d\Omega}\left(x',y';\theta_x=\frac{x-x'}{\Delta},\theta_y=\frac{y-y'}{\Delta}  \right),
\end{equation}
where the paraxial approximation has again been assumed in taking $\tan\theta_{x,y}\approx\theta_{x,y}$. Equations (\ref{eq:KramersMoyal1}) and (\ref{eq:KramersMoyal1b}) then have the simple interpretation that the intensity downstream of a sample, which is contained in the plane $z=0$, superposes a micro-structure-independent near-field Fresnel diffraction pattern that is damped with a decoherence factor\cite{Nesterets2008} $1-F$, with a diffusely-scattered signal $\mathcal{B}$ arising from the superposition of each SAXS fan emerging from each point over the exit surface of the sample.  Such a family of SAXS fans is measured e.g.~in Schaff et al.\cite{schaff2015}; this amounts to a measurement of $K(x,y,x',y',z=\Delta)$.   

In the form given by Eqs~(\ref{eq:KramersMoyal1}) and (\ref{eq:KramersMoyal1b}), the object-to-detector distance $\Delta$ is assumed to be sufficiently small that the detector is in the near-field of the slowly-varying part $\sqrt{I_s(x,y,z=0)}\exp[i\phi_{\textrm{s}}(x,y,z=0)]$ of the complex wave field at the sample exit surface $z=0$.  In these same equations,  $\Delta$ may be sufficiently large that the SAXS fan $K(x,y,x',y',z=\Delta)$---emerging from the point $(x',y',0)$ over the exit surface $z=0$ of the thin object, and considered as a function of coordinates $(x,y)$ over the detector in the plane $z=\Delta$---may have a width that is several pixels or more (cf.~Fig.~\ref{fig:appendix}(e)).  If these SAXS fans are highly structured, e.g.~if they exhibit SAXS satellite peaks from highly directional unresolved structures such as collagen fibres \cite{SAXScollagen}, then Eqs~(\ref{eq:KramersMoyal1}) and (\ref{eq:KramersMoyal1b}) can be used without any need for further approximation, to model image contrast in a regime where the detector plane is in the near-field of the resolved structure in the sample, but in the far-field of the unresolved micro-structure.   

However if $\Delta$ is indeed small enough that the SAXS fan spreads over no more than a few pixels, and if the SAXS fans are not too highly structured, an intermediate simplification is possible, with a domain of validity between that of the less-general Fokker--Planck equation and the more general expression in  Eqs~(\ref{eq:KramersMoyal1}) and (\ref{eq:KramersMoyal1b}). This intermediate model arises from a more precise consideration of the angular scattering distribution in the local SAXS differential cross sections, compared to that given in the preceding section.  With this end in mind, assume that the angular widths $\theta(x',y')$ of all differential scattering cross sections contained in $K(x',y',x,y,z=\Delta)$ obey $\theta(x',y')\ll 1$, with all such narrow SAXS fans being strongly peaked in the forward direction.  If this is the case, we are justified in Taylor expanding $I_s(x',y',z=0)$ about $(x,y)$, so that Eq.~(\ref{eq:KramersMoyal1b}) becomes: 
\begin{eqnarray}\label{eq:KramersMoyal4}
\mathcal{B}(x,y,z \!\!\!\!\!\! &=&  \!\!\!\!\!\!  \Delta) \approx F(x,y) \\ \nonumber  &\times& \!\!\!\!\!\!\!\!\!  \left[ I_s(x,y,z=0) + \sum_{m=0}^2 D_{m,2-m}^{(2)}(x,y; \Delta) \frac{\partial ^2 I_s(x,y,z=0)}{\partial x^m \partial y^{2-m}} + \sum_{m=0}^3 D_{m,3-m}^{(3)}(x,y;\Delta) \frac{\partial ^3 I_s(x,y,z=0)}{\partial x^m \partial y^{3-m}} + \cdots \right] . 
\end{eqnarray}
In obtaining the above expression, we have made use of the following three points:

\begin{itemize}

    \item We have assumed that each SAXS fan $K (x,y,x',y',z=\Delta)$---emanating from the point $(x',y')$ over the exit surface of the sample to produce an intensity distribution that is a function of detector coordinates $(x,y)$ in the plane $z=\Delta$---obeys:
\begin{equation}
\iint K (x,y,x',y',z=\Delta) \, dx' dy' =1.    
\end{equation}
This normalisation condition is appropriate since the multiplicative factor of $F$ is a scattering {\em fraction} which accounts for the fact that only a certain percentage of the incident beam is converted to SAXS upon passage through the sample.

    \item We have assumed that 
\begin{equation}\label{eq:SAXS-fan-centroid-is-at-the-origin}
\iint (x'-x) K (x,y,x',y',z=\Delta) \, dx' dy'  = \iint (y'-y) K (x,y,x',y',z=\Delta) \, dx' dy' =0, \quad m=1,2,    
\end{equation}
consistent with the fact that any transverse shift in the centroid of the SAXS fan may be interpreted as a local gradient in $\phi_{\textrm{s}}(x,y)$ that has already been accounted for in Eq.~(\ref{eq:KramersMoyal1}).  This eliminates the terms $D_{01}^{(1)}(x,y;\Delta)$ and $D_{10}^{(1)}(x,y;\Delta)$ that would otherwise have appeared in Eq.~(\ref{eq:KramersMoyal4}). There is an intrinsic ambiguity here: it is question of semantics whether we consider $\phi_{\textrm{s}}(x,y)$ to possess a linear phase gradient and assume the SAXS fan to have a centre-of-mass that is undeflected, or whether we instead consider there to be no local gradient in $\phi_{\textrm{s}}(x,y)$ but view the centre-of-mass of the SAXS fan to be transversely shifted on account of correlations that are present in the sample's unresolved micro-structure.  In writing Eq.~(\ref{eq:SAXS-fan-centroid-is-at-the-origin}) we take the former of these two equivalent points of view.  

    \item We have defined the following hierarchy of diffusion tensors (SAXS-fan moments, cf.~Gureyev et al.\cite{DeblurByDefocus}), which arise from terms of progressively higher order in the Taylor expansion of $I_s(x',y',z=0)$ about $(x,y)$:
\begin{eqnarray}\label{eq:KramersMoyal5}
D_{m,2-m}^{(2)}(x,y;\Delta) \!\!\!\!\!\! &=& \!\!\!\!\!\! \frac{1}{2!} \binom{2}{m} \iint (x'-x)^m(y'-y)^{2-m} \, K (x,y,x',y',z=\Delta) \, dx'dy', \quad\quad m=0,1,2,
\\ \label{eq:KramersMoyal5b} D_{m,3-m}^{(3)}(x,y;\Delta) \!\!\!\!\!\! &=& \!\!\!\!\!\! \frac{1}{3!} \binom{3}{m} \iint (x'-x)^m(y'-y)^{3-m} \, K (x,y,x',y',z=\Delta) \, dx'dy', \quad\quad m=0,1,2,3,\\ 
\nonumber \vdots
\\
\label{eq:KramersMoyal5c} D_{m,M-m}^{(M)}(x,y;\Delta) \!\!\!\!\!\! &=& \!\!\!\!\!\! \frac{1}{M!} \binom{M}{m} \iint (x'-x)^m(y'-y)^{M-m} \, K (x,y,x',y',z=\Delta) \, dx'dy', \quad\quad \!\!\!\! m=0,1,2,\cdots,M.
\\ \nonumber \vdots
\end{eqnarray}
Here, $\binom{M}{m}$ denotes a binomial coefficient $M!/[m!(M-m)!]$. The moments in Eq.~(\ref{eq:KramersMoyal5}) quantify the ellipticity of the SAXS fan, via the three independent diffusion-tensor components $D_{02}^{(2)}(x,y),D_{11}^{(2)}(x,y),D_{20}^{(2)}(x,y)$, which may be mapped to the semi-major axis of the SAXS ellipse, the semi-minor axis and the rotation angle\cite{CrankBook}. The higher-order diffusion tensors in Eqs~(\ref{eq:KramersMoyal5b}--\ref{eq:KramersMoyal5c}) have an analogous interpretation.

\end{itemize}

Finally, if we insert Eq.~(\ref{eq:KramersMoyal4}) into Eq.~(\ref{eq:KramersMoyal1}), we obtain the following finite-difference form of a Kramers--Moyal equation\cite{Risken1989}:
\begin{eqnarray}\label{eq:KramersMoyal6}
    \overline{I}(x,y,z=\Delta) \!\!\!\!\!\! &=& \!\!\!\!\!\! I_s(x,y,z=0) -(\Delta/k)\left[1-F(x,y)\right]   \nabla_{\perp}\cdot[I_s(x,y,z)\nabla_{\perp}\phi_{\textrm{s}}(x,y,z)]_{z=0}  \\  \nonumber &+& \!\!\!\!\!\! F(x,y) \, \left [\sum_{m=0}^2 D_{m,2-m}^{(2)}(x,y; \Delta) \frac{\partial ^2 I_s(x,y,z=0)}{\partial x^m \partial y^{2-m}} + \sum_{m=0}^3 D_{m,3-m}^{(3)}(x,y;\Delta) \frac{\partial ^3 I_s(x,y,z=0)}{\partial x^m \partial y^{3-m}} + \cdots \right].
\end{eqnarray}

The $F(x,y)=0$ case of Eq.~(\ref{eq:KramersMoyal6}) is the commonly-employed first-order finite-difference form of the transport-of-intensity equation\cite{Teague1983} (see Eq.~(\ref{eq:PropagationBasedPhaseContrast})).  The second line augments this equation to include SAXS-induced blurring, in a manner which takes the anisotropy of the local SAXS fans into account. Pawula\cite{Pawula1967} has shown that such Kramers--Moyal expansions may be either truncated at most to second or lower orders in spatial derivatives, or not truncated at all.  In our context, this amounts to working with one of five possibilities, with increasing order of generality: 

\begin{itemize}
  
  \item Keeping only the first line of Eq.~(\ref{eq:KramersMoyal6}) (TIE approximation), with $F=0$, which ignores SAXS altogether;
  
  \item Setting $D^{(3)}$ and all higher-order diffusion tensors to zero, and assuming rotationally-invariant SAXS fans via
\begin{equation}\label{eq:RotSymSAXSFan}
D_{02}^{(2)}(x,y;\Delta)=D_{20}^{(2)}(x,y;\Delta)\equiv D(x,y;\Delta), \quad D_{11}^{(2)}(x,y;\Delta)=0,    
\end{equation}
leaving us with the Fokker--Planck form given in Eq.~(\ref{eq:FPEin2D_finite_difference2}) if we can also assume that $F\ll 1$;

  \item Setting $D^{(3)}$ and all higher-order diffusion tensors to zero, thereby reducing Eq.~(\ref{eq:KramersMoyal6}) to an anisotropic Fokker--Planck equation incorporating the diffusion tensor $D_{mn}^{(2)}(x,y;\Delta)$  (assumption of an elliptic SAXS fan); 

  \item Keeping all orders in Eq.~(\ref{eq:KramersMoyal6}), which contains all SAXS-fan moments.   Note that this hierarchy of SAXS-fan moments may be measured, e.g.~by raster scanning a focused beam through a thin sample to obtain a family of SAXS patterns\cite{Fratzl1997,schaff2015} from which each component of each diffusion tensor at each location $(x,y)$ may be obtained using Eqs~(\ref{eq:KramersMoyal5})-(\ref{eq:KramersMoyal5c});
  
  \item Eschewing the Kramers--Moyal and Fokker--Planck approximations and instead working with Eqs~(\ref{eq:KramersMoyal1}) and (\ref{eq:KramersMoyal1b}) directly.
  
\end{itemize}

\section{Discussion}

The primary goal of this paper is to give several means by which the Fokker--Planck equation (and its extension to the Kramers--Moyal equation) may be used to model forward and inverse problems of near-field x-ray imaging, in the presence of both phase--amplitude shifts and non-negligible small-angle x-ray scattering from unresolved micro-structure in a thin sample.    

A detailed study of any particular application or applications is beyond the scope of this paper.  One such application, grating-based x-ray phase contrast imaging\cite{David,Momose,Weitkamp,Pfeiffer2008df}, is studied in detail in a companion paper\cite{MorganPaganin2019} that gives a number of analytical expressions based on the finite-difference expression in Eq.~(\ref{eq:FokkerPlanckXray}). More generally, Eqs~(\ref{eq:FokkerPlanckXray}) and (\ref{eq:FPEin2D}) may be used to calculate, in one and two transverse dimensions respectively, the near-field paraxial image corresponding to a thin sample with known position-dependent exit-surface intensity, phase and diffusion coefficient.  As we have already seen, a sufficiently slowly varying fraction $F(x,y)$---this being the fraction of the incident radiation that is converted to SAXS---may be brought into the second square brackets on the right hand side of Eqs~(\ref{eq:FPEin2D}) and (\ref{eq:FokkerPlanckXray}), and thereby incorporated into an effective diffusion coefficient.  See Eqs~(\ref{eq:EffectiveDiffusionCoefficient}),  (\ref{eq:RelationBetweenPhaseDepthandF}) and (\ref{eq:DiffusionCoeff2}), in addition to Eq.~(g) in Fig.~\ref{fig:RulesOfThumb}.  These expressions can be used for calculating propagation-based phase contrast x-ray images\cite{Snigirev,Cloetens,WilkinsFish} and speckle-tracking phase contrast images \cite{berujon2012,Morgan2012,zdora2018}, as well as the previously mentioned grating-based phase-contrast images.  Note, also, that if one is in the intermediate-field rather than the near-field regime for $\sqrt{I_{\textrm{s}}}\exp(i \phi_{\textrm{s}})$, expressions such as (i) Eqs~(\ref{eq:Appendix09}) and (\ref{eq:Appendix13}); or (ii) Eqs~(\ref{eq:Appendix09}) and (\ref{eq:KramersMoyal1b}); or (iii) Eqs~(\ref{eq:Appendix09}) and (\ref{eq:KramersMoyal4}) may instead be used.  All of these expressions are amenable to analytical implementation if the functional forms of intensity, phase and diffusion coefficients are known, or to computational implementation if the intensity, phase and diffusion are numerically modelled e.g.~using Geant4\cite{Paterno2018}.      

The above indications, of the forward problem of near-field and intermediate-field x-ray imaging for a thin sample in the presence of both coherent and diffusive scattering, set up Fokker--Planck and Kramers--Moyal equations that can be used to consider the associated inverse problem of reconstructing intensity, phase and diffusion from measured intensity data.  It is again beyond the scope of our paper to explore this point further, beyond the indicative suggestions given in the three examples below.

\bigskip

{\em Example \#1}. As a first example of inverse problems that may be tackled based on the formalism of this paper, observe that if the diffusion-tensor moments are measured e.g.~using the previously-mentioned technique of raster scanning a focused quasi-monochromatic x-ray probe over a lattice of points at the entrance surface of thin object\cite{Fratzl1997,schaff2015}, and the scattering fractions $F(x,y)$ obtained using the same data, then the bottom line of Eq.~(\ref{eq:KramersMoyal6}) will be known.  Since both $\overline{I}(x,y,z=\Delta)$ and $I_s(x,y,z=0)$ can be measured using quasi-monochromatic x rays of the same wavelength, and since both the object-to-detector distance $\Delta$ and the radiation wave number $k$ are known, the only remaining unknown in Eq.~(\ref{eq:KramersMoyal6}) would be the phase $\phi_{\textrm{s}}$.  This phase could then be obtained using existing techniques for numerically solving the transport-of-intensity equation (e.g.~a full multi-grid method\cite{Gureyev_1999} or a method using four Fourier transforms\cite{paganin1998}). 

\bigskip

{\em Example \#2}. A second application of the formalism of the present paper, is to the field of x-ray speckle tracking\cite{berujon2012,Morgan2012}.  The technique of x-ray speckle tracking illuminates an object with a known {\em resolved} reference speckle field, and then measures one or more distorted forms of that speckle field which arise when a sample is placed in the illuminating speckle beam. The diffuse-scatter signal, considered in the present paper, arises naturally in x-ray speckle tracking---via unresolved sample-induced speckles that should be carefully distinguished from the titular illuminating speckles---and gives useful information that may be extracted using a variety of techniques. These techniques are typically based on the reduction in illuminating-speckle visibility due to SAXS-induced diffusion: see e.g.~the review by Zdora\cite{zdora2018}, and references therein.    The geometric-flow approach to speckle tracking, which has recently been developed for x-rays\cite{PaganinLabrietBrunBerujon2018} but has also been very recently applied to visible-light imaging \cite{Lu2019}, takes as a starting-point the following equation:
\begin{equation}\label{eq:GeometricFlow1}
    I_R(x,y)-I_S(x,y)=\nabla_{\perp}\cdot[I_R(x,y) \, {\bf{D}}_{\perp}(x,y)].
\end{equation}
This relates the intensity $I_R(x,y)$ of a reference speckle image obtained in the absence of a sample, to the distorted/coded/encrypted form of that speckle image, denoted $I_S(x,y)$, obtained in the presence of a pure-phase-object sample; ${\bf{D}}_{\perp}(x,y)$ is a displacement-vector field, which distorts the reference speckle image $I_R(x,y)$ into the speckle image $I_S(x,y)$ obtained in the presence of the sample. The above equation, modelled on the continuity equation that has played such a dominant role in the present paper, conserves photon energy both globally and locally.  It may be used to reconstruct the speckle displacement ${\bf{D}}_{\perp}(x,y)$ induced by the phase object, and hence a map of the phase shift $\phi(x,y)$ due to the object, in a simple and efficient manner that implicitly rather than explicitly tracks speckles\cite{PaganinLabrietBrunBerujon2018}.  Note that the displacement field is related to the phase shift of the object, via
\begin{equation}
  {\bf{D}}_{\perp}(x,y)=(\Delta/k)\nabla_{\perp}\phi(x,y),    
\end{equation}
where $\Delta$ is the sample-to-detector distance and $k$ is the wave-number of the x-rays. A Fokker--Planck form of Eq.~(\ref{eq:GeometricFlow1}), suitable for a phase object described by both its phase shift $\phi(x,y)$ and effective diffusion coefficient $D_{\textrm{eff}}(x,y;\Delta)$, is:
\begin{equation}\label{eq:GeometricFlow2}
    I_R(x,y)-I_S(x,y)=\nabla_{\perp}\cdot[I_R(x,y) \, {\bf{D}}_{\perp}(x,y)] - \Delta \nabla_{\perp}^2[D_{\textrm{eff}}(x,y;\Delta) I_R(x,y)].
\end{equation}
This incorporates both coherent flow and diffuse flow, into the geometric-flow method for x-ray speckle tracking.  It would be interesting to investigate an augmented form of the geometric-flow x-ray speckle-tracking method, which takes Eq.~(\ref{eq:GeometricFlow2}) rather than Eq.~(\ref{eq:GeometricFlow1}) as a starting point, and seeks to reconstruct both $\phi(x,y)$ and $D_{\textrm{eff}}(x,y;\Delta)$ rather than just $\phi(x,y)$.  Of relevance in such a context---namely, of separating out the signal due to coherent versus diffusive energy flow---is the fact that $D_{\textrm{eff}}(x,y;\Delta)$ will transform as a scalar upon rotation of the sample through 180 degrees (about a rotation axis that is perpendicular to the optic axis, e.g.~in a computed-tomography setting), whereas ${\bf{D}}_{\perp}(x,y)$ will transform as a vector under the same rotation.

\bigskip

{\em Example \#3}.  As a third and somewhat more detailed example of inverse problems that may be tackled using the formalism outlined in the present paper, again consider a static non-magnetic non-crystalline sample that is made of a single material.  This sample may and in general will have varying spatial density, with complex refractive index\cite{Paganin2006}
\begin{equation}
n(x,y,z)=1-\delta(x,y,z)+i\beta(x,y,z)    
\end{equation}
which is such that the ratio $\delta(x,y,z)/\beta(x,y,z)$ is independent of position at all points within the sample\cite{paganin2002,paganin2004}.  The associated unresolved micro-structure is left arbitrary, and the exit surface of the sample is assumed to correspond to the plane $z=0$. Further assume normally-incident unit-intensity illumination by $z$-directed quasi-monochromatic x-ray plane waves.  This single-material approximation, together with the projection approximation\cite{Paganin2006}, implies that the x rays at the exit surface of the sample have an intensity $I(x,y,z=0)$ and phase $\phi(x,y,z=0)$ given by
\begin{equation}
I(x,y,z=0)=\exp[-\mu T(x,y)],\quad \phi(x,y,z=0)=-k\delta T(x,y).    
\end{equation}
Here $\mu=2 k \beta$ is the linear attenuation coefficient of the material from which the sample is composed, and $T(x,y)$ is the projected thickness of the sample along the $z$ direction.  The fact that\cite{paganin2002}
\begin{equation}
\nabla_{\perp}\cdot\left[I(x,y,z=0)\nabla_{\perp}\phi(x,y,z=0))\right]=\nabla_{\perp}\cdot\left\{ \exp[-\mu T(x,y)] \nabla_{\perp} [-k \delta T(x,y)] \right\}=\frac{k\delta}{\mu}\nabla_{\perp}^2 \exp[-\mu T(x,y)]
\end{equation}
enables the 2+1-dimensional finite-difference Fokker--Planck equation in Eq.~(\ref{eq:FPEin2D_finite_difference2}) to be transformed to:
\begin{eqnarray}\label{eq:InverseProblem1}
I(x,y,z=\Delta)=\left\{ 1 - \Delta\left[
\frac{\delta}{\mu} -  D_{\textrm{eff}}(x,y,\Delta)  \right]\nabla_{\perp}^2\right\} \exp[-\mu T(x,y)], \quad F(x,y) \ll 1.
\end{eqnarray}
Here, we have used Eq.~(\ref{eq:EffectiveDiffusionCoefficient}) to write the effective diffusion coefficient $D_{\textrm{eff}}(x,y,\Delta)$. Equation~(\ref{eq:InverseProblem1}) is a linear elliptic partial differential equation for the unknown exponentiated projected thickness $\exp[-\mu T(x,y)]$ of the single-material sample with projected thickness $T(x,y)$, and is an algebraic equation for the unknown effective diffusion coefficient (SAXS term, or so-called ``dark-field'' signal\cite{MorrisonBrowne1992,SuzukiUchida1995,Pfeiffer2008df}) $D_{\textrm{eff}}(x,y,\Delta)$. Note that the $D_{\textrm{eff}}(x,y,\Delta)\rightarrow 0$ limit of Eq.~(\ref{eq:InverseProblem1}) is the basis of a commonly-used single-image method for x-ray phase retrieval\cite{paganin2002}.  When we cannot assume $D_{\textrm{eff}}(x,y,\Delta)$ to vanish, we may solve Eq.~(\ref{eq:InverseProblem1}) for both $D_{\textrm{eff}}(x,y,\Delta)$ and $T(x,y)$, as follows.  Suppose propagation-based phase contrast images are obtained at two different propagation distances, $\Delta_1$ and $\Delta_2$.  From Eq.~(f) in  Fig.~\ref{fig:RulesOfThumb}, we can write down the scaling relation:
\begin{equation}\label{eq:InverseProblem3}
    \frac{D_{\textrm{eff}}(x,y,\Delta_1)}{\Delta_1}=\frac{D_{\textrm{eff}}(x,y,\Delta_2)}{\Delta_2}.
\end{equation}
This scaling results from the already-mentioned fact that the slab of vacuum, in between the exit surface $z=0$ of the object and the entrance surface $z=\Delta$ of the detector, is not a diffusive medium.  Thus the width of the SAXS fans scales linearly with $\Delta$, rather than being proportional to $\Delta^{1/2}$ (as would be the case for true diffusion). Setting $D\propto\Delta$ ensures that the final term on the right side of Eq.~(\ref{eq:DiffusionEqnFiniteDifferenceForm}) is proportional to $\Delta^2$, ensuring the width of the SAXS fan scales as $(\Delta^{1/2})^2=\Delta$, as required.  As is the case throughout the paper, the artifice of a $\Delta$-dependent diffusion coefficient is a price to be paid for the simplicity of being able to work {\em simultaneously} with both diffusion concepts and free-space coherent propagation concepts, in the context of the Fokker--Planck and Kramers--Moyal equations for x-ray imaging. To proceed we write the $z=\Delta_1$ and the $z=\Delta_2$ cases of Eq.~(\ref{eq:InverseProblem1}), and then use   Eq.~(\ref{eq:InverseProblem3}) to eliminate $D_{\textrm{eff}}(x,y,\Delta_{1,2})$.  Hence:
\begin{equation}\label{eq:InverseProblem4}
\frac{\Delta_2^2 \, I(x,y,z=\Delta_1)-\Delta_1^2 \, I(x,y,z=\Delta_2)}{\Delta_2^2-\Delta_1^2}=\left( 1-\frac{\delta}{\mu} \frac{\Delta_1\Delta_2}{\Delta_1+\Delta_2} \nabla_{\perp}^2\right) \exp[-\mu T(x,y)].
\end{equation}
Equation~(\ref{eq:InverseProblem4}) is mathematically identical in form to Eq.~(7) of Paganin et al.\cite{paganin2002}, and may therefore be solved using the same Fourier-transform method.  This gives:
\begin{equation}\label{eq:InverseProblem5}
T(x,y)=-\frac{1}{\mu}\log_{\textrm{e}}\left\{\mathcal{F}^{-1}\left[\frac{\mathcal{F}
\left[ \Delta_2^2 \, I(x,y,z=\Delta_1)-\Delta_1^2 \, I(x,y,z=\Delta_2)\right]     
}{1+\frac{\delta}{\mu} \frac{\Delta_1\Delta_2}{\Delta_1+\Delta_2}(k_x^2+k_y^2)}\right]\right\}+\frac{1}{\mu}\log_{\textrm{e}}(\Delta_2^2-\Delta_1^2), \quad \Delta_1\ne \Delta_2.
\end{equation}
Here, $\mathcal{F}$ denotes Fourier transformation with respect to $x$ and $y$, in any convention that transforms $\partial/\partial x$ into $ik_x$ and $\partial/\partial y$ into $i k_y$; $(k_x,k_y)$ are Fourier coordinates corresponding to $(x,y)$, and $\mathcal{F}^{-1}$ is the inverse of $\mathcal{F}$. Equation~(\ref{eq:InverseProblem5}) shows how a linear combination of two images, taken at two different distances $\Delta_1$ and $\Delta_2$, may be combined to yield the projected thickness $T(x,y)$ of a thin single-material sample, in a manner that is independent of SAXS scattering by the sample (cf.~Pagot et al.\cite{Pagot2003}).  Once $T(x,y)$ has been so obtained, Eq.~(\ref{eq:InverseProblem1}) can then be solved algebraically for $D_{\textrm{eff}}(x,y)$.

\bigskip

In the third example above, and indeed throughout the paper, the effective diffusion coefficient (and, more generally, the diffusion tensors as defined in Eqs~(\ref{eq:KramersMoyal5}--\ref{eq:KramersMoyal5c})) will have at least three independent contributions:
\begin{itemize}
    \item The first contributor to the hierarchy of diffusion tensors is the sample-induced scatter due to unresolved micro-structure, that has been a main theme of the present paper.  
    \item The second contributor to the diffusion tensors is an edge-scattering signal\cite{MorrisonBrowne1992,SuzukiUchida1995} that is now known to be due to unresolved sharp edges\cite{Yashiro2015}, and which may be viewed as a Young--Maggi--Rubinowicz type boundary wave\cite{YoungOnTheBoundaryWave,Maggi,Rubinowicz,MiyamotoWolf1,MiyamotoWolf2,BornWolf} or a Keller-type\cite{Keller} diffracted ray.  This is examined in further detail in the previously mentioned companion paper\cite{MorganPaganin2019}.  
    \item The third contribution to the diffusion tensors is associated with incoherent aberrations (including the modulation transfer function of the detector, and penumbral blurring effects due to finite source size) of the imaging system that is used to measure images of the sample.  As an example: take Eq.~(\ref{eq:InverseProblem1}), set $D_{\textrm{eff}}(x,y,\Delta)$ to a constant $D_{\textrm{eff}}(\Delta)$ that is independent of $x,y$ so as to model penumbral blur due to finite source size in a linear shift-invariant imaging system, and then note that the resulting equation has a solution identical to that recently published by Beltran et al.\cite{Beltran2018} 
\end{itemize}
All three effects will in general be present simultaneously, so that e.g.~the SAXS fans associated with sample-induced blurring will need to be convolved with the intensity point-spread function of a given optical imaging system, and similarly with the signal associated with unresolved sharp edges.  These interesting complications---which may be summarised in the statement that ``partial coherence can come from the source, or the object itself''---are beyond the scope of the present paper.

There is some relation between the differential form of the diffusion operators employed in this paper, and the idea of Laplacian-based unsharp-mask image sharpening\cite{Unsharp1,Unsharp2}.  From an unsharp-mask image processing perspective, the operator\cite{Unsharp0} 
\begin{equation}
    \mathcal{L}=1-w^2(x,y)\nabla_{\perp}^2 
\end{equation}
locally sharpens an image $I(x,y)$, with a characteristic position-dependent partial-deconvolution transverse length scale equal to $w(x,y)$; cf.~Eq.~(\ref{eq:InverseProblem1}).  The approximate inverse of this operator, which locally blurs an image over the same transverse length scale $w(x,y)$, is (cf.~Eqs~(\ref{eq:DiffusionEqnFiniteDifferenceForm}) and (\ref{eq:BlurOperatorSecondOrder}))\cite{Unsharp0}: \begin{equation}
   \mathcal{L}^{-1}\approx 1+w^2(x,y)\nabla_{\perp}^2. 
\end{equation}
This blurring is rotationally symmetric.  If this rotational symmetry is not present, we can use a Kramers--Moyal type operator to effect local blurring with a position-dependent point-spread function, namely
 \begin{equation}
     \mathcal{K}=1+\sum_{m=0}^2 w_{m,2-m}^{(2)}(x,y) \partial_{x}^m \partial_{y}^{2-m} + \sum_{m=0}^3 w_{m,3-m}^{(3)}(x,y) \partial_{x}^m \partial_{y}^{3-m} + \cdots.
\end{equation}
The notation above is analogous to the second line of Eq.~(\ref{eq:KramersMoyal6}), and we have set $\partial/\partial x \equiv\partial_{x},\partial/\partial y \equiv\partial_{y}$.  This gives another perspective on the use of derivative operators to effect blurring associated with a position-dependent point spread function.  Similar ideas are used in image sharpening via coherent defocus\cite{DeblurByDefocus} and image blurring operations of the convolution type\cite{Gureyev2017Unreasonable}.

The two derivations of the Fokker--Planck equation for paraxial imaging, given in the present paper, may be augmented by other approaches.  An approach based on Wigner functions\cite{Alonso2011} would be an interesting avenue for future research, since Wigner functions naturally capture the idea that there can be a distribution of energy flow vectors at each point in space \cite{Chandrasekhar1943}. See e.g.~Eq.~(19) of Nugent and Paganin\cite{NugentPaganin2000}, which was obtained via a Wigner-function formulation of paraxial optics, and which may be readily manipulated into a Fokker--Planck form given by Eq.~(\ref{eq:FPEin2D_finite_difference2}).  From a different perspective again, the intensity transport equations associated with arbitrary linear shift-invariant optical imaging systems\cite{PaganinPRA2018}, may be rendered into Fokker--Planck and Kramers--Moyal forms via the slight generalisation of replacing shift-invariant local differential operators with their corresponding shift-variant forms, along very similar lines to those indicated in the preceding paragraph.

Unresolved speckle lies at the heart of many phenomena observed in imaging using partially coherent optical fields \cite{Paganin2006,PaganinDelRio2019}.  Here, ``speckle'' is used in the general sense of optical fields whose spatial and/or temporal intensity fluctuations have a random component; this differs from the more usual usage which equates ``speckle'' with ``fully developed speckle''. Unresolved speckles have two distinct origins: (i) the partial coherence of the illuminating beam and its associated spatio-temporal speckles, and (ii) unresolved speckle associated with sample micro-structure.    In the the present paper, the spatial width $w$ of the point-spread function (PSF) of the position-sensitive detector---a lower limit on which is given by the pixel size---defines a natural transverse length scale.  Speckles much smaller in size than $w$ will be spatially averaged when registering an image.  If ergodicity of the associated stochastic process can be assumed, over the area of each pixel, then by definition the spatially-averaged intensity at each pixel location can be replaced with a corresponding ensemble average over many realisations of the ensemble of fields illuminating each pixel.  Physically, and as has been used in the present paper, this corresponds to a statistical ensemble of objects, each of which has the same coarse-grained projected complex refractive index distribution, but differing spatially-random micro-structure.  This construct permits us to approximate the resulting measured intensity map, with a probability distribution obeying the 2+1-dimensional Kramers--Moyal equation, including the special case of the 2+1-dimensional Fokker--Planck equation.  Refraction and prism terms are associated with the resolved coarse-grained structure of the ensemble of objects and the ensemble of illuminations, with the diffusion term being associated with the unresolved speckles. What is deemed SAXS is a detector-resolution-dependent construct associated with unresolved sample-induced speckles.  Thus, the component of the scattered radiation attributed to SAXS is dependent on $w$, and thereby detector dependent.  This last statement ties in with the broader idea that there is a close link between partial coherence and the spatio-temporal resolution of one's detector---see e.g.~Sec.~3.6.3 of Paganin\cite{Paganin2006}.  It is also closely related to what component of the x-ray energy flow is classed as ``coherent energy transport'', and what is classed as ``diffusive energy transport'', in the Fokker--Planck and Kramer--Moyal constructions: the finer the spatio-temporal resolution of the intensity detection, the smaller the fraction of the x-ray energy flow that will be classed as diffusive.   Both the TIE-type and SAXS-type scattering (resolved and unresolved structures) arise from the same fundamental physics, as is clear in the first-principles derivation in Sec.~2.2. Again, what length scales ultimately fall into either category is determined by the resolution of the detector.

We close this discussion by emphasising the broader applicability, beyond the margins delineated in the present paper, of the Fokker--Planck and Kramers--Moyal equations to (i) partially coherent paraxial x-ray imaging in particular, and (ii) partially-coherent paraxial imaging more generally. Context may be given to this claim, by the huge domain of applicability of Fokker--Planck and Kramers--Moyal concepts, as cited in Sec.~1. Of equal relevance is the likelihood that the domain of validity of the paraxial-imaging Fokker--Planck and Kramers--Moyal equations is at least as broad at the transport-of-intensity equation which they generalise; and, since the transport-of-intensity equation has been applied to visible light \cite{Barty1998}, electrons\cite{Bajt2000}, neutrons \cite{Allman2000} and x-rays \cite{paganin2002}, there is every expectation that the formalism of the present paper may be utilised for paraxial-imaging scenarios using all of the just-listed radiation and matter wave fields.  Extensions of the present paper, to {\em imaging} applications beyond those listed previously, include a Fokker--Planck approach to neutron phase contrast imaging\cite{KleinOpat1976,Allman2000}, x-ray magnetic circular dichroism imaging\cite{Eimuller2001}, x-ray edge-illumination imaging \cite{Olivo2001}, x-ray Zernike phase contrast\cite{Neuhausler2003}, neutron grating imaging \cite{pfeiffer2006neutron}, electron out-of-focus contrast imaging (Fresnel contrast imaging)\cite{CowleyBook} and visible-light out-of-focus contrast imaging \cite{Barty1998}.  As with the present study, a core motivation for such an approach is its potential simultaneously to model, under the aegis of single powerful formalism, both deterministic and stochastic aspects of paraxial thin-object imaging using partially coherent radiation and matter wave fields.    

\section{Conclusion}

Fokker--Planck and Kramers--Moyal equations were obtained, to model near-field and intermediate-field paraxial quasi-monochromatic x-ray imaging of thin samples exhibiting both phase--amplitude and small-angle-scattering contrast.  Two  complementary derivations were given.  Possible applications, to both forward and inverse imaging problems, were discussed.

\bibliography{sample}

\section*{Acknowledgements}  We acknowledge useful discussions with Mario Beltran, Jeremy Brown, Tim Davis, Carsten Detlefs, Timur Gureyev, Alexander Kozlov, Kieran Larkin, Thomas Leatham, Heyang (Thomas) Li, Andrew Martin, Konstantin Pavlov, Tim Petersen and Florian Schaff. We acknowledge the following funding sources: ARC Future Fellowship FT180100374; Veski Victorian Postdoctoral Research Fellowship (VPRF); German Excellence Initiative and European Union Seventh Framework Program (291763).

\section*{Author contributions statement}

D.M.P. and K.S.M. worked on this paper in close collaboration. D.M.P. prepared all figures and performed all mathematical calculations, arising from discussions between both authors.  The paper was mainly written by D.M.P., with input from K.S.M. 

\section*{Additional information}


{\bf Competing financial interests:} The authors declare no competing financial interests. 

\end{document}